# Impedance spectroscopy, dielectric and magnetic study of M-type Strontium hexaferrite ($SrFe_{12}O_{19}$) and its modified systems [$SrFe_{11.7}Co_{0.45}O_{19}$, $SrFe_{11.7}Ti_{0.225}O_{19}$, $SrFe_{11.7}Co_{0.225}Ti_{0.1125}O_{19}$]

Ranjit Pattanayak[1*]

[1]Department of Physics and Astronomy, National Institute of Technology, Rourkela-769008, India

*ranjit.p20@gmail.com

Tel.: +91 661 2462723, Fax. +91 661 2462999

**Abstract**

Polycrystalline M-type hexagonal Strontium hexaferrite ($SrFe_{12}O_{19}$) and its modified systems [$SrFe_{11.7}Co_{0.45}O_{19}$, $SrFe_{11.7}Ti_{0.225}O_{19}$, $SrFe_{11.7}Co_{0.225}Ti_{0.1125}O_{19}$] were successfully prepared by the solid state reaction method. The single phase and well grain growth of micrometer range was confirmed by XRD pattern and SEM image respectively. The impedance spectroscopy and dielectric properties were investigated in the frequency range of 100 Hz–1 MHz & temperature range of 30 ºC-200 ºC. The room temperature M-H loop study for $SrFe_{12}O_{19}$ and $SrFe_{11.7}Co_{0.45}O_{19}$ was carried out by VSM. From the investigation it was found that, the microstructure has (grain, grain boundary and electrode polarization) significant effect on dielectric relaxation and conduction mechanism for all the systems. From M-H loop study it was found that Co modified system has drastic reduction of coercive field.

## 1. Introduction

Hexaferrites are classified into four types depending on chemical and crystalline structure. These include hexaferrites types M, W, Y and Z. The crystalline and magnetic structures of the different types of hexaferrites are remarkably complex, a part of that M-type hexagonal ferrites, $MFe_{12}O_{19}$(M=Ba, Sr, Pb), with the magnetoplumbite structure continue to be important permanent magnet materials in microwave, small motor, and, more recently magnetic recording applications. The unit cell of M-hexaferrite contains ten oxygen layers, sequentially constructed for four blocks: S (spinel), R (hexagonal), S* and R*. The S* and R* blocks have equivalent atomic arrangements, but rotated at 180º with respect to S and R blocks around the c-axis. An S or S* block consists of two $O^{2-}$ layers; while R or R* block contains three $O^{2-}$ layers, with one oxygen site in the middle layer substituted by an $Ba^{2+}/Sr^{2+}/Pb^{2+}$. M-type hexagonal ferrites crystallizes in a hexagonal structure with 64 ions per unit cell on 11 sites of different symmetry. atoms are distributed over five distinct sites: three octahedral sites (12k, 2a and 4f2), one tetrahedral site (4f1) and one bipyramidal site (2b). The magnetic structure given by the Gorter model is ferrimagnetic with five different sub-lattices, three parallel (4f1 and 4f2) which are coupled by super exchange interactions through the O ions. The magnetic properties of M-type ferrites can be tailored by the partial substitution of $Ba^{2+}/Sr^{2+}/Pb^{2+}$ or $Fe^{3+}$ ions, or both by magnetic or non magnetic cations[1,2]. Among M-type hexagonal ferrite, Strontium hexaferrites (SrM) have been widely investigated and used as permanent magnets because of their high magnetization($M_s$), Curie temperature ($T_c$) and coercivity (H), environmental stability, and low price for production. SrM is also used to minimize the electromagnetic interference (EMI)[3,4][5].Many number of groups has been synthesized SrM by using various techniques (solid state, wet-chemical processes like sol–gel citrate, hydrothermal method, micro-emulsion process) and studied the physical properties review[1][6,7]. While, another group of researchers have concentrated their studies on substitution of divalent-tetravalent ions at iron site and observed avariation in magnetic properties of strontium ferrite (SrM) [8–15]. Although large number of investigation have been carried out to study, analyze and improve the magnetic properties of SrM system, the high temperature electrical properties with the role of microstructures (grain, grain boundary and electrode surface) of this system have not been studied up to the same extent.

*Want et. al.*[16]have been synthesized Lanthanum substitution strontium hexaferrite (SrM) and studied their dielectric, conducting and impedance related studies as a function frequency ranges of 20 Hz–3 MHz at room temperature (RT). From impedance they have reported bulk and grain boundary have significant effect on overall dielectric response. *Ashiq et. al. [17]*have reported about nanosized strontium hexaferrite doped with a binary mixture of Al–Cr at the iron site that, the variation in the dielectric constant and dielectric loss factor with frequency was follwed the basis of Wagner and Koop's theory. *Again Ashiq et. al.[18]* have reported Zirconium copper substituted calcium strontium hexagonal ferrites shown variable conductivity with doping. Likewise Iqbal et. al. synthesized Calcium substituted SrMand they reported, the DC electrical resistivity increases with increase of calcium content up to some level and then decreased on further addition of calcium where the variations of dielectric constant and dielectric loss angle were followed the Maxwell–Wagner and Koops models.

Considering the above previous reports the present work devote towards the investigating the electric transport properties of SrM as well as substitution of Co, Ti and co-substitution of Co-Ti systems by the help of Complex Impedance Spectroscopy technique within temperature and frequency range of 30ºC to 200ºC and frequency range 100Hz to 1MHz.

## 2. Experimental Technique

Polycrystalline Strontium ferrite ($SrFe_{12}O_{19}$) (SrM) and substitution by $Co^{2+}$, $Ti^{4+}$ ($SrFe_{12-x}M_xO_{19}$ : x=0.45 for Co substitution, x=0.225 for Ti substitution) and co-substitution of Co-Ti ($SrFe_{12-x}Co_xTi_yO_{19}$ : x=0.225 and x=0.1125)  at iron site, was prepared by the conventional ceramic method. To keep the net valance of SrM system constant, the variable value 'x' was considered. In this way we have kept the iron as Fe: $Fe_{11.7}$ and the added number of different cations were approximately constant. For synthesis initially, Iron oxide ($Fe_2O_3$), strontium carbonate ($SrCO_3$), cobalt oxide (CoO), titanium oxide ($TiO_2$) powder were considered as raw materials.

These powders were mixed together according to their molecular weight ratios and grinded for 2 h and calcinated at 1100 °C for 3 h. The calcinated powder again grinded for 1 h mixed with binder (PVA) then pressed into cylindrical pellets and sintered at 1200 °C for 5 h. The pure phase of sintered pellet was verified by X-ray diffraction pattern. The surface morphology was studied by JEOL JSM-6480LScanning Electron Microscopy system. The dielectric parameters were measured using a computer-controlled impedance analyzer (HIOKI IMPEDANCE ANALYZER 1352) as a function of frequency (100 Hz–1 MHz) in the temperature range between 30 ºC (RT) and 200 ºC. The room temperature M–H loop of the samples was obtained using vibrating sample magnetometer (VSM).

## 3. Results and discussion:

### 3.1 Structural and microstructural study

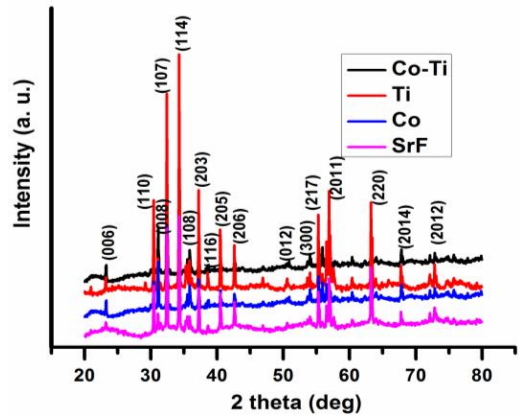

**Fig. 1. X-ray diffraction patterns of SrFe$_{12}$O$_{19}$ and modified systems.**

Fig. 1 shows the XRD patterns corresponds to pure phase formation of SrF and Co, Ti & Co-Ti co-substitution at Fe site of SrF system. The position of observed peaks was verified with corresponding JCPDS pattern.

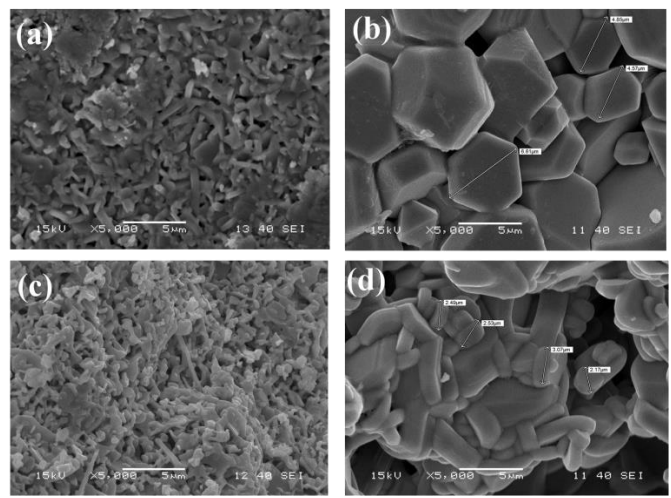

**Fig. 2. SEM micrographs of SrFe$_{12}$O$_{19}$ (SrM) and modified systems (a) SrM (b) Co (c) Ti & (d) Co-Ti.**

The SEM micrograph (Fig. 2a, 2b, 2c and 2d) illustrates the dense grains of SrF and Co, Ti& Co-Ti co-substituted systems. From SEM micrograph it was observed that, grain size increases significantly with Co substitution.

### 3.2 Impedance and Modulus Study

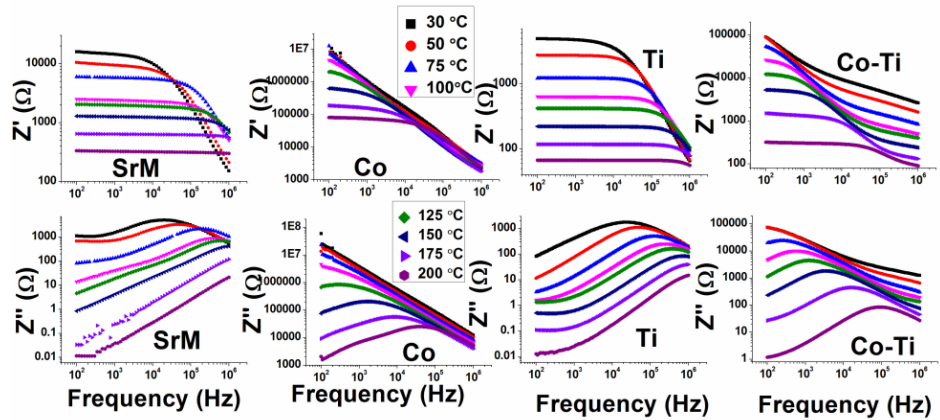

**Fig. 3. Frequency dependence of (a) real part (Z') and (b) imaginary part (Z'') ofimpedance at different temperatures for SrM and its modified systems.**

The Fig. 3 shows the variation of resistive [Z'(ω)] and reactive part [Z'' (ω)] of the systems within the frequency and temperature domain of 100 Hz to 1 MHz & 30°C (room temperature) to 200°C respectively. The real and imaginary impedance was calculated by using the formula as: Z'=Zcosθ, Z''=Zsinθ, where Z=(Z'+Z''), i=√-1 is the imaginary factor [19][20]. The reduction of Z'(ω) (having single plateau region connected with a dispersive region at higher frequency side) with temperature is the evidence of semiconducting nature of all the studied system. But from variation of imaginary impedance (Z''(ω)) it was found that, except Co substituted system, all the systems have single relaxation peak corresponding to the Z' plateau region. For Co substituted system, with increasing temperature, around 150°C another plateau region enters through low frequency window. Basically, the real (Z') and imaginary (Z'') part of the impedance significantly depends upon grain and grain boundary resistance (R) and capacitance (C) [21]:

$$Z' = \frac{R_g}{(1+R_g\omega_g C_g)^2} + \frac{R_{gb}}{(1+R_{gb}\omega_{gb} C_{gb})^2} \qquad (1)$$

$$Z'' = \frac{-R_g^2 \omega_g C_g}{1+(R_g\omega_g C_g)^2} + \frac{-R_{gb}^2 \omega_{gb} C_{gb}}{1+(R_{gb}\omega_{gb} C_{gb})^2} \qquad (2)$$

Where $R_g$, $C_g$, $R_{bg}$ and $C_{gb}$ are grain and grain boundary capacitance & resistance.

It was well known that for a polycrystalline system each microstructure (grains, grain boundaries and other interfaces) has its distinctive conduction mechanism which activates at different temperatures and frequency ranges [22]. The complex impedance spectroscopy can segregate the grain, grain boundary and interfaces relaxation/conduction mechanism (due to different relaxation times) effect by separating semi-circles in complex impedance plane (cole-cole of impedance i.e. Z' vs. Z" plot). The resistance for the grain and grain boundary can be calculate from the intercepts on the real part (Z') of impedance, whereas the capacitance values have been calculated from the frequency peaks of the semi-circle arcs [20,23].

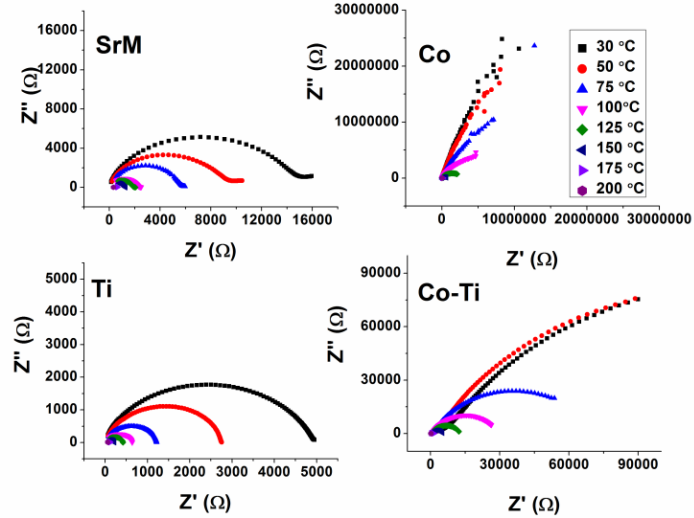

**Fig. 4. . Cole-Cole of impedance at different temperatures of SrM and its modified systems.**

The Fig. 4 shows the cole–cole plot [$Z'(\omega)$ Vs. $Z''(\omega)$] of all the studied system. It was observed that, all the systems have single semicircle at RT, however, the radius of semicircle (resistance) continuously decreases with increasing temperature (for all the systems). It was also found that, Co substituted system has shown largest resistance whereas Ti substituted system has shown smallest resistance as compared with others. It was well known that the single relaxation peak in Z"/single semicircle in cole-cole of impedance may has double conduction mechanism due to presence of co-contribution effect which can be verify from cole-cole of modulus (M' Vs, M")[24][20]. The real and imaginary part of modulus were calculated by using the formula as: M'=$C_o$Z"and M"=$\omega C_o$Z'where M=M'+iM"[23][25]. Here, $C_o=\varepsilon_o A/t$ is the vacuum capacitance, $\varepsilon_o=8.85*10^{-12}$ F/m is the permittivity of the free space, t is the thickness of the sample and A is the cross sectional area of the electrode deposited on the sample and $\omega$ is the angular frequency.

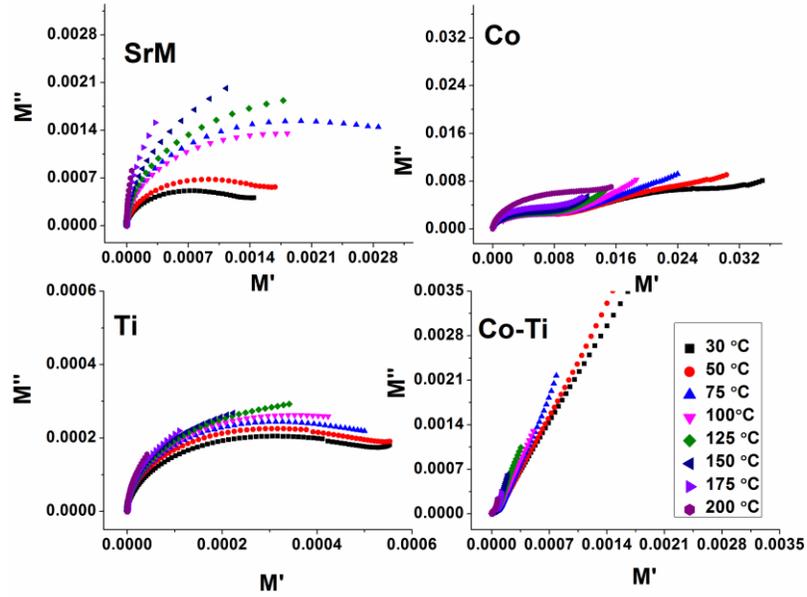

**Fig. 5. Cole-Cole of modulus at different temperatures for SrF and its modified systems.**

The Fig. 5 shows the cole–cole modulus plot [M′ Vs. M″] of all the studied systems with increasing temperature. It was found that (Fg. 5), for SrM, Ti substituted SrM and Co-Ti co-substituted SrM have single semicircles in the studied temperature range which was matched with cole-cole of impedance spectra. However, for Co substituted SrM has shown double relaxation behavior even at room temperature (RT), with increasing of temperature both the radius of semicircles decreases. Similar type of behavior was observed for bulk $BaFe_{12}O_{19}$ (BaM) system, i.e. grain and grain boundary have co-contribution effect which was segregated by modulus study[20]. Here it was confirmed that, the appeared single semicircle for all the systems except Co substituted system is due to grain boundary effect but extra the semicircle appeared in cole-cole of modulus (which was absent in cole-cole of impedance) plot at higher frequency side is due to grain contribution(for Co substitution system). So it was confirmed that Co substitution system has co-contribution of grain and grain boundary effect at RT, however, other systems have only grain boundary effect the evolution of grain effect for Co substituted system may be due to increasing of grain size.

## 3.3. Dielectric analysis

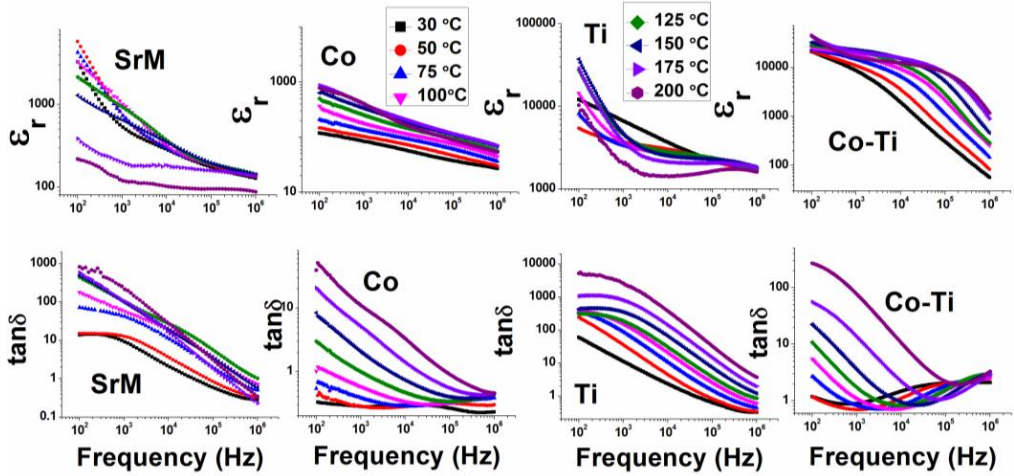

**Fig. 6. Frequency dependence of dielectric constant ($\varepsilon_r$) and dielectric loss (tan$\delta$) at different temperatures for SrM and its modified systems at different temperatures**

The Fig. 6 represents the variation of dielectric constant ($\varepsilon_r = Ct/\varepsilon_0 A$) with frequency at different temperatures for SrM and other modified systems. It was found that (Fig. 6), Ti substituted system has large dialectic constant with high loss than other systems. Likewise, Co substituted system has least dielectric constant with low loss than other systems. The giant dielectric constant with high loss can explain on the basis of Maxwell-type interfacial polarization [26]. According to Maxwell-Wagner type interfacial polarization concept the real ($\varepsilon^{'}$) and imaginary ($\varepsilon^{''}$) permittivities were formularized as [26,27]:

$$\varepsilon^{'}(\omega) = \frac{1}{C_0(R_i+R_b)} \frac{\tau_i+\tau_b-\tau+\omega^2\tau_i\tau_b\tau}{1+\omega^2\tau^2} (3)$$

$$\varepsilon^{''}(\omega) = \frac{1}{\omega C_0(R_i+R_b)} \frac{1-\omega^2\tau_i\tau_b+\omega^2\tau(\tau_i+\tau_b)}{1+\omega^2\tau^2} \quad (4)$$

where subindex *i* and *b* refer to the interfacial-like and bulk like layers, respectively,

$$\tau_i = C_i R_i, \tau_b = C_b R_b, \tau = \frac{(\tau_i R_b + \tau_b R_i)}{R_i + R_b}, C_0 = \frac{\varepsilon_0 A}{t}$$

*A*=area of the capacitor, *t* =thickness, R=Resistance and C=Capacitance

From equ. (3) and (4) it can be conclude that, if the resistive nature of grain or grain boundary will be change it has great impact on the dielectric constant as well as loss for a system due to dropping of the applied electric field mostly charge-depleted interfacial [28]. The prepared hexaferrite systems have larger number of $Fe^{3+}$ and $Fe^{2+}$ ions so the observed polarization is due to exchange of electrons takes place between $Fe^{3+}$ and $Fe^{2+}$ ions. For Ti substitution system $Ti^{4+}$ cations were gone to the Fe sites which increases the net free charge carriers at interfaces as the result, the system has shown largest dielectric constant even at RT. Likewise, for Co substitution system $Co^{2+}$ cations were gone to the Fe sites which can deceases the net free charge carriers which shown least dielectric constant. Therefore the dielectric constant value for mixed substitution system ($Co^{2+}$-$Ti^{4+}$) was varies in between Ti and Co substitution system.

Fig. 6 represents the variation of dielectric loss (tanδ) of all the systems with frequency at various temperatures. It was well known that, for a polycrystalline system, the dielectric loss develops due to lags behind the applied ac electric field which causes by impurities and imperfections or defects in the materials [29]. From our SEM study it can be observe that Co-substitution system has larger grain size i.e. least grain boundary density (defects), so the increasing of grain size (reduction of grain boundary density) is may be a another cause for reduction of dielectric loss for Co substituted SrM system as compared with others [30][31].

## 3.4. Conductivity Study

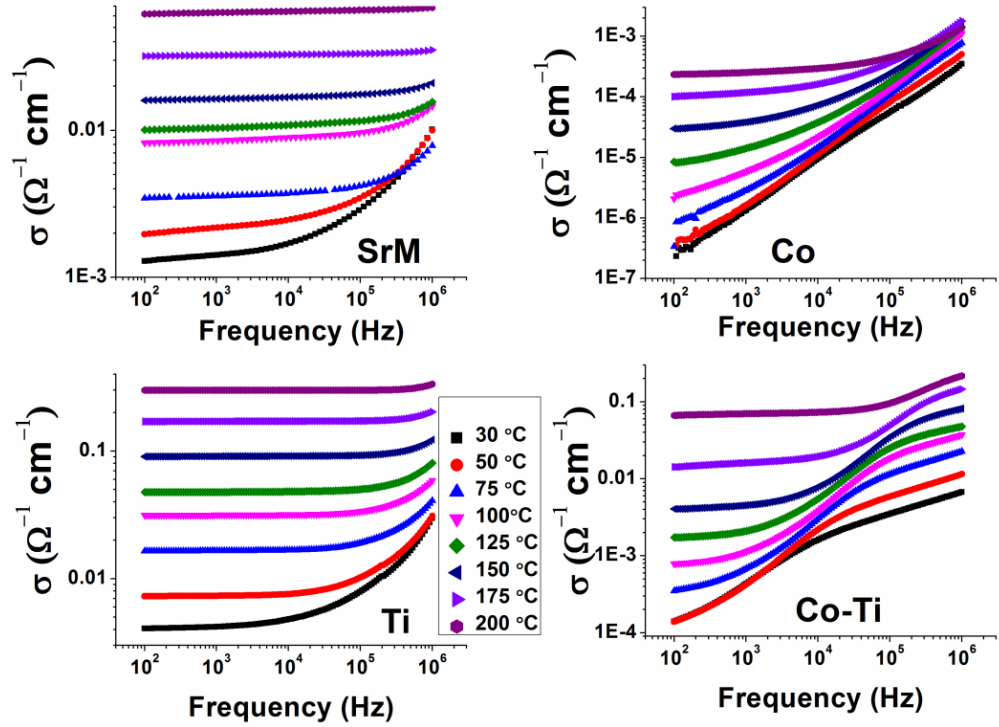

**Fig. 7. Frequency dependence of ac Conductivity for SrM and its modified systems at different temperatures.**

The Fig. 7. shows the variation of ac conductivity with frequency at different temperatures of SrM and modified systems, which were derived from the formula $\sigma = [Z'/(Z'^2 + Z''^2)] (t/A)$ [32].

At RT (Fig 7), Co-substituted system, a frequency independent region (plateau regions) appears which smoothly relaxes to a strong dispersive region towards higher frequency for SrM and other systems which is indicating the existing of single potentials countered by charge carriers near grain boundaries [33].

From Fig. 7. It was also found that, Ti-substituted system has higher conductivity at RT as compared with other system whereas Co-substituted system has least one. The behavior of all the observed conductivity spectra can be explain on the base of jumping relaxation model [34].

According to this model, two types of hopping process can possible such as : (a) successful hopping and (b) unsuccessful hopping.

In case of successful hopping, the activated ions can easily follow the applied frequency as the result they hops to a new site permanently which gives dc conductivity (plateau regions). In case of unsuccessful hopping, the ions couldn't able to follow the applied frequency as result the permanent hopping process hundreds which gives ac conductivity (dispersive regions). From ac conductivity spectra (Fig. 7) it was found that, at RT, Ti substituted system has shown longest frequency range for dc conduction whereas Co substituted system has least. This variable dc conduction is may be due to variable oxidation states of substituted cations.

To know the detail hopping mechanism the conductivity spectra can be fit by using the single power law given by Jonscher[35] (which will be consider in the future work).

$$\sigma(\omega) = \sigma(0) + A\omega^n \quad (5)$$

Here, $\sigma(\omega)$ is the total conductivity, $\sigma(0)$ corresponding to dc conductivity and $A\omega^n$ characterized to frequency dependent conductivity known as ac conductivity. Where, $\omega$ is applied ac frequency. The 'n' (frequency dependent exponent) and 'A' are the temperature and material intrinsic property dependent constants.

### 3.5. Magnetic Study

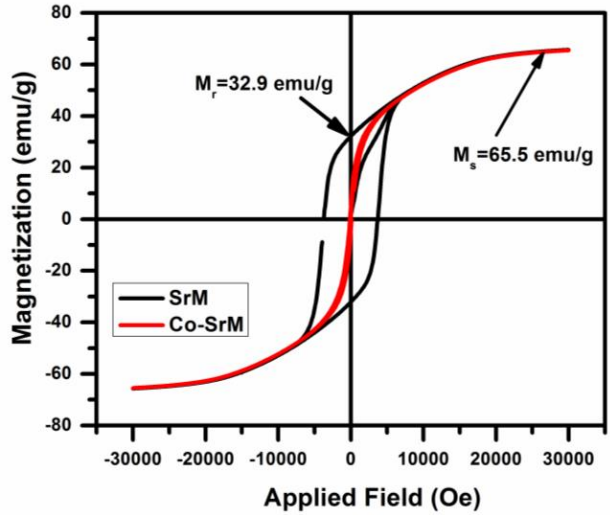

**Fig. 8. Room temperature M–H loop of SrM and Co-substituted SrM systems.**

The M–H loops at room temperature for SrM and Co substituted SrM are shown in Fig. 8. From the figure it was observed that both the system exhibits well saturated hysteresis loop having ferromagnetic nature. It was also observed that, the saturation magnetization ($M_s$) for both the systems are same however Co substituted SrM has shown drastic reduction of coercive field ($H_c$) as like soft magnetic behavior. Similar type of reduction of coercive field has been found for single crystalline of SrM and Mg-Ti substitution of BaM [12,23,36] . Recently we have verified that the grain-grain interfaces are key factor to tune the coercive field. Therefore it was concluded that both increasing of grain size and $Co^{2+}$ substitution at iron sites are the cause behind the drastic reduction of coercive field.

### 4. Conclusion

Polycrystalline Strontium hexaferrite ($SrFe_{12}O_{19}$) andits modified systems [$SrFe_{11.7}Co_{0.45}O_{19}$, $SrFe_{11.7}Ti_{0.225}O_{19}$, $SrFe_{11.7}Co_{0.225}Ti_{0.1125}O_{19}$] were successfully prepared by the solid state reaction technique. From SEM micrograph it was found that Co substitution helps increasing of grain size. From impedance and modulus spectroscopy it was observed that, Co substitution system has co-contribution of grain and grain boundary effect where as other systems have only grain boundary effect. From dielectric study it was observed that Co substituted system has shown smallest dielectric constant and loss with highest resistance. Whereas, Ti substituted system has shown highest dielectric constant and high loss with smallest resistance. The dielectric behavior of all the systems were well obeyed the Maxwell-Winger model. From M-H loop study it was found that Co modified system has shown drastic reduction of coercive field as compared with SrM system.

## Acknowledgment

The authors want to acknowledge Prof. D. Behera, Department of Physics and Astronomy, NIT Rourkela for valuable discussions. To NiharikaMohapatra, Indian Institute of Technology, Bhubaneswar, Odisha, for collection of magnetic data and Ministry of Human Resource Development (MHRD), Government of India for providing scholarship.